\documentclass[twocolumn,showpacs,preprintnumbers,amsmath,amssymb,eps,prb]{revtex4}
                           
\usepackage{graphicx}
\usepackage{dcolumn}
\usepackage{bm}

\begin{document}

i

\title{Calculations for an inhomogeneous $d$-wave superconductor}

\author{E. S. Caixeiro}
\affiliation{Centro Brasileiro de Pesquisas F\'{\i}sicas, Rio de Janeiro, RJ 22290-180 Brazil\\
\label{1}}%
\author{E. V. L. de Mello}
\affiliation{
Departamento de F\'{\i}sica, Universidade Federal Fluminense, Niter\'oi, RJ 24210-340, Brazil\\}%

\author{A. Troper}
\affiliation{Centro Brasileiro de Pesquisas F\'{\i}sicas, Rio de Janeiro, RJ 22290-180 Brazil
}%
\affiliation{ Universidade do Estado do Rio de Janeiro, R. S\~ao Francisco 524,  Rio de Janeiro 20550013, Brazil\\
}%
\date{\today}

\begin{abstract}
We calculate the distributions $P[\Delta_0({\bf r}_{i})]$ of local $d$-wave pairing 
amplitude $\Delta_0({\bf r}_{i})$ at a position ${\bf r}_{i}$ inside a 
disordered high temperature 
superconductor (HTS) family. To reproduce the observed 
inhomogeneity a random potential $V^{imp}$, within a Bogoliubov-de Gennes (BdG) formalism, 
is considered. We perform calculations with fixed values of the disorder strength 
$V^{imp}$ obtaining the distribution of $\Delta_0({\bf r}_{i})$, and local density 
of charge carriers $\rho({\bf r}_{i})$, for different compounds of the LSCO family. 
The calculation of the relative root-mean-square deviation shows that the underdoped 
compounds are more inhomogeneous than the overdoped ones. Also, the spatial variation 
of $\Delta_0({\bf r}_{i})$ decreases as the average density of charge carriers  
$\langle \rho \rangle$ increases, demonstrating that the compounds are more homogeneous 
for high values of $\langle \rho \rangle$. 
The results indicate that the $d$-wave superconducting gaps seem to be more sensitive 
to a change in the disorder in comparison with the $s$-wave superconducting gaps.

\end{abstract}

\maketitle

\section{Introduction}
It is clear by now that HTS have many non-conventional  physical properties,
and they are the reason why the fundamental interaction is yet to be discovered. In some cases it is 
possible  that these unusual properties result from the variation in the average density of 
holes (or charge carriers) $\langle \rho \rangle$ in the CuO$_{2}$ planes~\cite{Muller,Lee,Evandro3}.

Many experimental features show that the density of charge carriers $\langle \rho \rangle$ 
plays an important role in the physics of the HTS. It is observed that the zero 
temperature superconducting gap $\Delta_0$ increases when $\langle \rho \rangle$ 
diminishes~\cite{Moura,Harris,Egami,Oda}, which is an unexpected behavior since 
the critical temperature $T_c$ vanishes for low values of 
$\langle \rho \rangle$~\cite{Caixeiro0}. It is also experimentally observed 
that for some compounds the electrical charges are highly inhomogeneous in the
CuO$_{2}$ planes~\cite{Egami,Buzin00,Fournier}. These intrinsic charge 
inhomogeneities~\cite{Egami2} are not related to crystal defects. Although it 
may be not so strong  in some cuprate superconductors~\cite{Bobroff,LoramI,Caixeiro3,Evandro2}. 
The origin of the inhomogeneity observed in these materials may be related 
to the fact that the holes injected into the CuO$_{2}$ planes by chemical 
doping may leave behind charged dopant ions~\cite{Wang}. Such inhomogeneity may produce 
regions with spatially varying $\Delta_0({\bf r}_i)$ and a variation in the local 
density of states (LDOS)~\cite{Pan}. Many similar results from scanning
tunneling microscopy were reported lately on clean surfaces\cite{McElroy,Vershinin}.
Inside the bulk, Nuclear quadrupolar resonance (NQR)
experiments\cite{Singer} have also measured that the charge inhomogeneity
increases as the temperature decreases, exactly as one expects in a 
phase separation transition\cite{Muller,Evandro3}. 

Since an inhomogeneous medium  does not have translational symmetry, we applied 
the Bogoliubov-deGennes (BdG) formalism~\cite{Ghosal,
Soininen,Franz1,Franz2,Evandro3,Wang}, conceived originally to deal with finite 
systems. To reproduce the intrinsic charge 
inhomogeneity we consider a nonmagnetic local disorder potential, which has only 
the effect of changing the local chemical potential on each site of the 
lattice~\cite{Ghosal}. Our purpose here is to obtain a distribution of local 
superconducting gaps $\Delta_0({\bf r}_{i})$, and a distribution of local density 
of charge carriers $\rho({\bf r}_{i})$ for an entire HTS family. We show that with 
the same magnitude of the disorder strength, the underdoped compounds become 
more inhomogeneous than the overdoped ones, which is in accordance with neutron 
diffraction experiments~\cite{Buzin00}. For the $d$-wave gap we observe that 
for very low disorder strength the distribution of local superconducting gaps 
gains significant weight near $\Delta_0({\bf r}_i)$=0 and the distribution of 
local charge carriers starts to bifurcate. This behavior differs from the 
$s$-wave gap symmetry distributions~\cite{Ghosal}, which needs higher values 
of disorder strength to have similar behavior.

\section{The Hamiltonian}

To describe the charge carriers dynamics in the CuO$_2$ planes of the HTS we consider 
an extended Hubbard Hamiltonian in a square lattice
\begin{eqnarray}
H&=&-\sum_{ \{ ij \} \sigma }t_{ij}c_{i\sigma}^\dag c_{j\sigma}
+\sum_{i\sigma}(V_i^{imp}-\mu)n_{i\sigma}
\nonumber \\
&&
+U\sum_{i}n_{i\uparrow}n_{i\downarrow}+
\frac V 2 \sum_{\langle ij \rangle \sigma
\sigma^{\prime}}n_{i\sigma}n_{j\sigma^{\prime}},
\label{a}
\end{eqnarray}
where $c_{i\sigma}^\dag (c_{i\sigma})$ is the usual fermionic creation (annihilation) 
operator at site ${\bf r}_i$, with lattice parameter $a$=1 and spin 
$\sigma \lbrace\uparrow\downarrow\rbrace$. $n_{i\sigma} =  c_{i\sigma}^\dag c_{i\sigma}$ 
is the density operator, and $t_{ij}$ is the hopping between sites $i$ and $j$. 
$U$ is the magnitude of the on-site repulsion, and $V$ is the magnitude 
of the nearest-neighbour attractive interaction. $\mu$ is the chemical potential, 
and $V_i^{imp}$ is the magnitude of the local disorder potential at site $i$, 
defined here as a random independent variable, uniformly distributed over 
$[-V^{imp},V^{imp}]$, similar to what Ghosal et al~\cite{Ghosal} made to
a S-wave superconductor. The  value of $V_i^{imp}$ assigned to the site ${\bf r}_i$ 
controls the strength of the disorder. 
We do not know the origin of $V_i^{imp}$, but it is possible to speculate its origin: 
It can be due to phase segregation, as in the case of La$_2$CuO$_{4+\delta}$,  
which has been observed to form oxygen rich and poor phases~\cite{Greniu,Jor}, 
or ion diffusion as observed in microcrystals of Bi2212~\cite{Truccato}. 
These findings, together with the NQR experiment of Singer et al\cite{Singer},
led us argue that is very likely that the upper pseudogap line is  
the onset of phase separation or bimodal decomposition~\cite{Evandro3} in
Cuprates superconductors. However, independently 
of its origin, the effect of $V_i^{imp}$ enters in our calculations by the 
change of  the local chemical potential at each site $i$ and, as a consequence, 
the local density of charge carriers $\rho({\bf r}_i)$, gives rises to
the charge inhomogeneity. 
Whether such charge non-uniformity is in form of stripes\cite{Tranquada} is
still an open question.
There are many ways of introducing the effects of disorder, for instance, 
Ovchinnikov et al~\cite{Ovchinnikov} considered magnetic impurities and 
Nunner et al.~\cite{Nunner} assumed a disorder in the pairing strength 
coupled with $V_i^{imp}$.

\section{The local equations}
To apply the BdG theory to the Hamiltonian (\ref{a}), one may define 
the pairing amplitudes, or pair potentials~\cite{Franz1}
\begin{equation}
\Delta_{\delta}({\bf r}_i)=V \langle  c_{i\downarrow}c_{i+\delta\uparrow} \rangle,
\hspace{0.5cm}
\Delta_U({\bf r}_i)=U\langle c_{i\downarrow}c_{i\uparrow} \rangle,
\label{b}
\end{equation}
where ${\mbox{$\delta$}}$=$\pm$$\hat{\bf x}$,$\pm$$\hat{\bf y}$ are unit vectors for a square lattice. 
In the mean-field theory Eq.(\ref{a}) can be solved using these pairing 
amplitudes. The resulting effective Hamiltonian $H_{eff}$ is given by 
\begin{eqnarray}
H_{eff}&=&-\sum_{i\delta\sigma}t_{i,i+\delta}c_{i\sigma}^\dag c_{i+\delta\sigma}
+\sum_{i\sigma}(V_i^{imp}-\tilde \mu_i)n_{i\sigma}
\nonumber \\
&&
+\sum_{i\delta }[\Delta_{\delta}^*({\bf r}_i) c_{i\downarrow}c_{i+\delta\uparrow}
+ \Delta_{\delta}({\bf r}_i) c_{i+\delta\uparrow}^\dag c_{i\downarrow}^\dag]
\nonumber \\
&&
+\sum_{i}[\Delta_U({\bf r}_i)c_{i\uparrow}^\dag c_{i\downarrow}^\dag
+\Delta_U^*({\bf r}_i)c_{i\downarrow}c_{i\uparrow}],
\label{c}
\end{eqnarray}
where $\tilde \mu_i$=$\mu$-${U/2}\langle n_i\rangle$ is the local 
chemical potential, which incorporates the site dependent Hartree 
shift ${\frac U 2}\langle n_i\rangle$~\cite{Ghosal}. Both, the Hartree shift 
and $V_i^{imp}$ take care of the charge inhomogeneity of the system.
The electronic density is given by $\langle n_i\rangle$=$\Sigma_{\sigma}\langle n_{i\sigma}\rangle$, 
and the hole density is $\rho({\bf r}_i)$=1-$\langle n_i\rangle$. The hole 
type density of charge carriers of a specific compound is given by 
\begin{equation}
\langle \rho \rangle={\frac 1 N_s}\sum_{i=1}^{N_s}\rho({\bf r}_i),
\label{d}
\end{equation}
where $N_s$ is the number of sites of the $N\times N$ square lattice. Therefore, 
Eq.(\ref{d}) fixes the chemical potential $\mu$. The $H_{eff}$ is diagonalized 
by the BdG transformations~\cite{Ghosal}
\begin{eqnarray}
c_{i\uparrow}=&
\sum_n[\gamma_{n\uparrow}u_n({\bf r}_i)-\gamma_{n\downarrow}^{\dag}v_n^*({\bf r}_i)],
\nonumber \\
\nonumber \\
c_{i\downarrow}=&
\sum_n[\gamma_{n\downarrow}u_n({\bf r}_i)+\gamma_{n\uparrow}^{\dag}v_n^*({\bf r}_i)],
\label{e}
\end{eqnarray}
where $\gamma_{n\sigma}$ and $\gamma_{n\sigma}^{\dag}$ are quasiparticle operators. 
$u_n({\bf r}_i)$ and $v_n({\bf r}_i)$ are normalized amplitudes for 
each ${\bf r}_i$, and are obtained from the BdG equations~\cite{Ghosal, Franz1}
\begin{equation}
\begin{pmatrix} K         &      \Delta  \cr\cr
           \Delta^*    &       -K^*
\end{pmatrix}
\begin{pmatrix} u_n({\bf r}_i)      \cr\cr
                v_n({\bf r}_i)
\end{pmatrix}=E_n
\begin{pmatrix} u_n({\bf r}_i)       \cr\cr
                 v_n({\bf r}_i)
\end{pmatrix}
\label{f}
\end{equation}
with 
\begin{eqnarray}
Ku_n({\bf r}_i)&=&-\sum_{\delta}t_{i,i+\delta}u_n({\bf r}_i+{\delta})
+(V_i^{imp}-\tilde \mu_i)u_n({\bf r}_i)
\nonumber \\
\Delta u_n({\bf r}_i)&=&\sum_{\delta}\Delta_{\delta}({\bf r}_i)u_n({\bf r}_i+{\delta})
+\Delta_U({\bf r}_i)u_n({\bf r}_i),
\label{g}
\end{eqnarray}
and similar  equations for $v_n({\bf r}_i)$. These equations give the quasiparticle 
eigenenergies $E_n(\ge 0)$. The temperature dependent BdG equations are solved self-consistently 
together with the pairing amplitudes~\cite{Franz1,Franz2} 
\begin{eqnarray}
\Delta_U({\bf r}_i)=&-U\sum_{n}u_n({\bf r}_i)v_n^*({\bf r}_i)\tanh{\frac {E_n}{2k_BT}}
\label{h}
\\
\nonumber\\
\Delta_{\delta}({\bf r}_i)=&-{\frac V 2}\sum_n[u_n({\bf r}_i)v_n^*({\bf r}_i+{\mbox{$\delta$}})
\nonumber \\
&
+v_n^*({\bf r}_i)u_n({\bf r}_i+\mbox{$\delta$})]\tanh{\frac {E_n} {2k_BT}}.
\label{i}
\end{eqnarray}
\begin{figure}[ht]
\begin{center}
  \begin{minipage}[b]{.1\textwidth}
    \begin{center}
    \centerline{ \includegraphics[width=8.5cm]{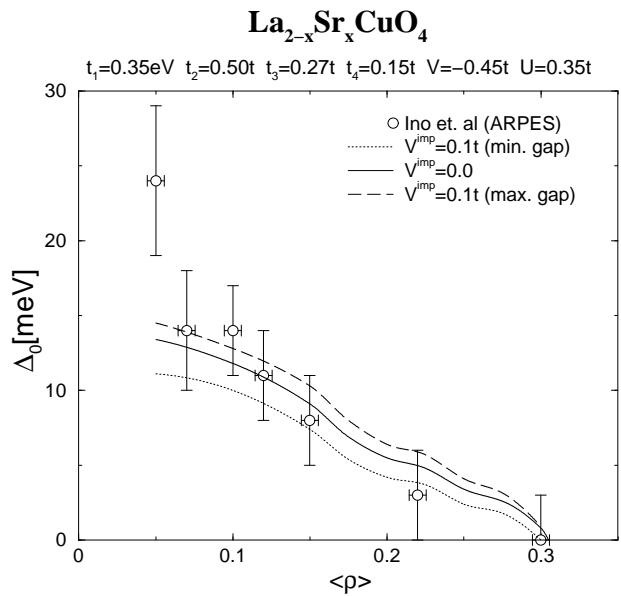}}
    \end{center} 
  \end{minipage}
\caption{The experimental leading edge gap of Ino et al.~\cite{Ino} (circles), 
interpreted as a superconducting gap, and the $\Delta_0$ curve as a function of 
the average density $\langle \rho \rangle$ for the La$_{2-x}$Sr$_{x}$CuO$_{4}$ 
family with $d$-wave symmetry. The solid line is the result without 
disorder ($V^{imp}$=0). The long-dashed and doted lines correspond to maximum 
and minimum pairing amplitudes, respectively, for the disorder strength $V^{imp}$=0.1$t$.}
\label{Fig0}
 \end{center}
\end{figure}
Eq.(\ref{i}) has four different possibilities of directions for a square 
lattice: $\Delta_{\pm \bf \hat x}({\bf r}_i)$ and $\Delta_{\pm \bf \hat y}({\bf r}_i)$. 
The combination of these terms may give rise to a $d$-wave gap or $s$-wave 
gap~\cite{Franz1}. The $d$-wave case is given by 
\begin{eqnarray}
\Delta_0({\bf r}_i)=&{\frac 1 4}[\Delta_{\bf\hat x}({\bf r}_i)+\Delta_{-\bf\hat x}({\bf r}_i)
-\Delta_{\bf \hat y}({\bf r}_i)-\Delta_{-\bf\hat y}({\bf r}_i)].\nonumber\\ 
\label{j}
\end{eqnarray}
Also, the local hole density of charge carriers is given by
\begin{eqnarray}
\rho({\bf r}_i)&=&1-2\sum_n[|u_n({\bf r}_i)|^2f_n+|v_n({\bf r}_i)|^2(1-f_n)],\nonumber\\ 
\label{k}
\end{eqnarray}
where $f_n$ is the Fermi function. With Eq.(\ref{k}) one can calculate $\rho({\bf r}_i)$ 
which, together with Eq.(\ref{d}), gives the density of charge 
carriers $\langle \rho \rangle$ of a compound.

Therefore, the BdG equations are solved self consistently, together 
with Eq.(\ref{h}) for $\Delta_U({\bf r}_i)$, Eq.(\ref{j}) for $\Delta_0({\bf r}_i)$, 
and with Eq.(\ref{k}) for $\rho({\bf r}_i)$, with periodic boundary conditions 
on a lattice with $N_s$ sites. We have performed calculations with lattices from 
$14\times 14$ to $24\times 24$, but here we concentrate on the parameters that 
reproduce the experimental results in a $22\times 22$.

\section{The local distributions}
\begin{figure}[ht]
\begin{center}
  \begin{minipage}[b]{.1\textwidth}
    \begin{center}
    \centerline{ \includegraphics[width=8.5cm]{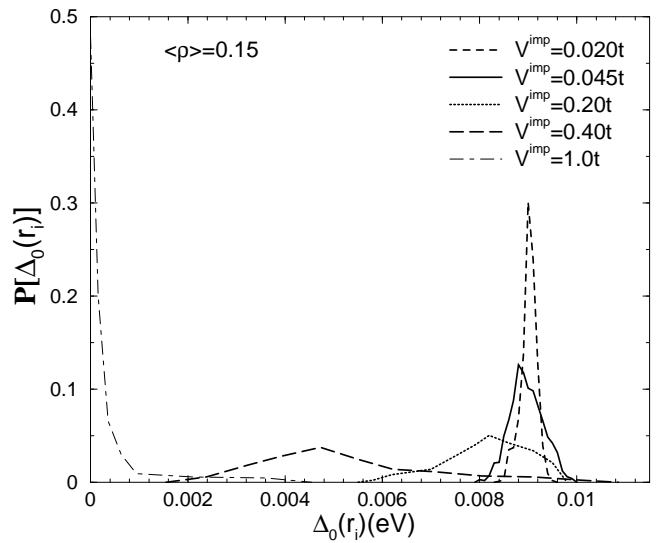}}
    \end{center} 
  \end{minipage}
\caption{The distribution $P[\Delta_0({\bf r}_i)]$ of local $d$-wave 
superconducting gaps $\Delta_0({\bf r}_i)$ for the near optimum compound $\langle \rho \rangle$=0.15, 
for different values of the disorder strength $V^{imp}$. At low 
disorder, $P[\Delta({\bf r}_i)]$ is peaked around the average gap value $\Delta_0$=9meV. 
As the disorder increases, $P[\Delta({\bf r}_i)]$ becomes broad. At 
very large disorder $P[\Delta({\bf r}_i)]$ gains significant weight near $\Delta_0({\bf r}_i)$=0.}
\label{Fig2}
 \end{center}
\end{figure}
\begin{figure}[ht]
\begin{center}
  \begin{minipage}[b]{.1\textwidth}
    \begin{center}
    \centerline{ \includegraphics[width=8.5cm]{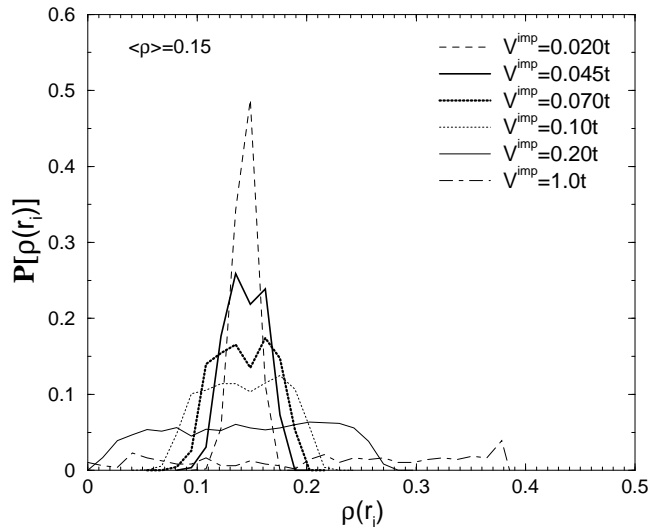}}
    \end{center} 
  \end{minipage}
\caption{The distribution $P[\rho({\bf r}_i)]$ of local density of charge 
for the near optimum compound, with average density value $\langle \rho \rangle$=0.15, 
for different values of the disorder strength $V^{imp}$. For low values of $V^{imp}$, 
$P[\rho({\bf r}_i)]$ is peaked around the average value. For $V^{imp}$=0.045$t$ 
and 0.070$t$, $P[\rho({\bf r}_i)]$ seems to have a bifurcation. For 
$V^{imp}$ greater than 0.1$t$, $P[\rho({\bf r}_i)]$ tends to spread.}
\label{Fig1}
 \end{center}                
\end{figure}

In order to reproduce the experimental phase diagram~\cite{Ino} 
$\Delta_0$ $vs$ $\langle \rho \rangle$ of  LSCO, we considered values 
of $U$ and $V$ close to values used in literature~\cite{Franz1,Franz2}. 
The constant coupling $U$ has low influence in the calculations of the 
superconducting gap for the $d$-wave symmetry~\cite{Angilella}. 
The hopping integrals were based on ARPES experiments results~\cite{Schabel} 
and we considered hopping integrals from the first to the fourth neighbour, 
differently from previous works with the BdG 
formalism~\cite{Ghosal,Soininen,Franz1,Franz2,Evandro3,Wang}. Here, 
the average superconducting gap is written as $\Delta_0$, 
different from the local superconducting gap, $\Delta_0({\bf r}_i)$.
The $\Delta_0({\bf r}_i)$ gap can be also obtained by a different
procedure through the study of the local density of states (LDOS)\cite{Daniel},
which is given by $N_i(E)=\sum_n[|u_n({\bf
x}_i)|^2f_n^{'}(E-E_n)+|v_n({\bf x}_i)|^2f_n^{'}(E+E_n)]$, where
the prime is the derivative with respect to the argument and
the $E_n$ are calculated by the BdG equations (Eq.\ref{f}). These
local calculations are important to interpret the new STM results
which show differences in the LDOS at mesoscopic scale\cite{Pan,McElroy,Vershinin}.

In Fig.\ref{Fig0} we plot the results for the average $d$-wave 
gap $\Delta_0$, for a $22\times 22$ lattice with the values: 
$t$=$t_1$=0.35eV, $t_2$=0.50$t$, $t_3$=0.27$t$, $t_4$=0.15$t$, $U$=0.35$t$ and $V$=-0.45$t$. 
These values are very close to previous calculations and 
known to reproduce the curve of $T^*$~\cite{Caixeiro3,Evandro2}.  
From Fig.\ref{Fig0} we observe that $\Delta_0$ diminishes as the density of charge 
carriers $\langle \rho \rangle$ increases, which is a common behavior of the 
superconducting gap for most HTS~\cite{Harris,Oda,Renner}. 
As mentioned $T^*$ $vs$ $\langle \rho \rangle$~\cite{Oda,Caixeiro3,Evandro2} decreases 
from high values for low densities (underdoped region), into low values for 
high  densities (overdoped region). Since experimental results~\cite{Harris,Oda,Ino} 
and theoretical calculations~\cite{Caixeiro3,Evandro2} indicate that 
the superconducting gap $\Delta_0$ of the HTS scales with $T^*$ and not with $T_c$, 
it is reasonable to expect a similar behavior of  $T^*$ $vs$ $\langle \rho \rangle$ 
and $\Delta_0$ $vs$ $\langle \rho \rangle$.  
The oscillations of the theoretical curves in Fig.\ref{Fig0} are associated 
with the number of hopping integrals: To embrace the entire range where 
the superconductivity of LSCO exists, it was necessary to consider hopping 
integrals from the first to the fourth neighbour, which
generates the oscillations in the $\Delta_0$ $vs$ $\langle \rho \rangle$ phase diagram.  
 
To study the disorder effects, with the same coupling parameters and hopping integrals 
used in the homogeneous case, we calculate the distributions of local 
$d$-wave superconducting gaps $P[\Delta_0({\bf r}_i)]$ and local hole 
densities $P[\rho({\bf r}_i)]$. In Fig.\ref{Fig2} we plot $P[\Delta_0({\bf r}_i)]$ for 
the  $\langle\rho\rangle$=0.15 compound. We observe that, for low disorder, 
$P[\Delta_0({\bf r}_i)]$ has a sharp peak near $\Delta_0({\bf r}_i)$$\approx$$9$meV, 
which is the homogeneous result, in accordance with Fig.\ref{Fig0}. As the 
disorder increases up to 1.0$t$, $P[\Delta_0({\bf r}_i)]$ becomes broad, and the 
system becomes more inhomogeneous. At very large disorder $P[\Delta({\bf r}_i)]$ 
gains significant weight near $\Delta_0({\bf r}_i)$=0. This behavior is similar to the 
$s$-wave gap, although it is necessary a disorder of 2$t$ in a 12x12 lattice, 
and 3$t$ in a 24x24 lattice, to obtain the same result for the $s$-wave 
case~\cite{Ghosal,Ghosal2}. Therefore, the $d$-wave gap is more sensible to 
changes in the disorder than the $s$-wave gap.  As one observes from 
Fig.\ref{Fig2}, for $V^{imp}$=0.045$t$ the disorder generates a distribution 
of $\Delta_0({\bf r}_i)$ around the average gap value $\Delta_0$. As we have already 
discussed in Refs.~\cite{Caixeiro3,Evandro2}, the maximum local gap is measured 
in the tunneling experiments and it seems to increase with the level of inhomogeneity\cite{Moura}. 
From Fig.\ref{Fig2} we see that, indeed, the disorder spreads the distribution 
$P[\Delta({\bf r}_i)]$, what increases the maximum local superconducting gap. 
Since the experimental leading edge shift, interpreted as the magnitude of the 
superconducting gap~\cite{Ino}, increases dramatically in the underdoped region, 
we can speculate that this is possibly the result of the inhomogeneity in 
the CuO$_2$ planes of the underdoped compounds. One can see the effect of 
$V^{imp}$=0.1$t$ on the maximum local superconducting gap curve in Fig.\ref{Fig0}.

It is important to notice that, since the values of $V^{imp}_i$ are chosen 
randomly at each site between the values in the interval $[-V^{imp},V^{imp}]$, 
the results are averaged over 10 realizations of disorder. With this procedure we have 
reproduced the results of the $s$-wave gap in a two dimensional 12x12 lattice of Ref.~\cite{Ghosal2}.  

In Fig.\ref{Fig1} we plot $P[\rho({\bf r}_i)]$ for the $\langle \rho \rangle$=0.15. 
We observe that, at low disorder, $P[\rho({\bf r}_i)]$ is peaked around 
the average density $\langle \rho \rangle$=0.15, and for $V^{imp}$=0.045$t$ 
and 0.070$t$, $P[\rho({\bf r}_i)]$ seems to have a bifurcation. As the disorder 
increases, $P[\rho({\bf r}_i)]$ also spreads considerably.  

\section{The inhomogeneous medium}

\begin{figure}[ht]
\begin{center}
  \begin{minipage}[b]{.1\textwidth}
    \begin{center}
    \centerline{ \includegraphics[width=8.5cm]{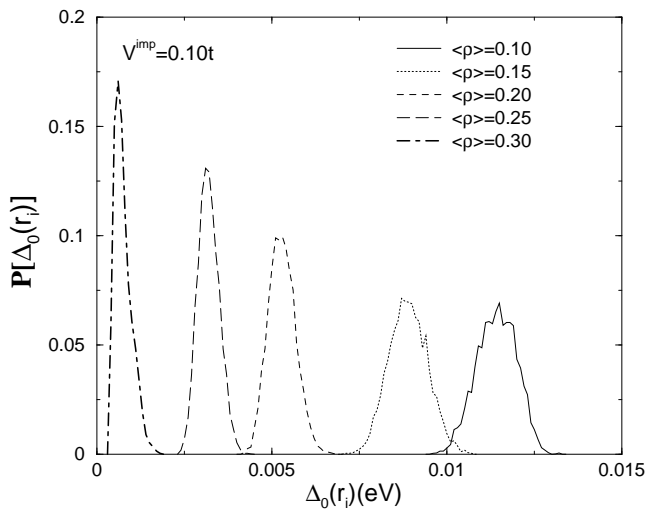}}
    \end{center} 
  \end{minipage}
\caption{The distribution $P[\Delta_0({\bf r}_i)]$ of local superconducting gaps for 
some selected compounds: the underdoped $\langle \rho \rangle$=0.10; the near 
optimum $\langle \rho \rangle$=0.15, and three overdoped compounds, 
$\langle \rho \rangle$=0.20, 0.25 and 0.30. The magnitude of the disorder 
is $V^{imp}$=0.10$t$. From the figure we observe that, as the average density 
$\langle \rho \rangle$ increases, the distributions $P[\Delta_0({\bf r}_i)]$ 
become sharp and thin.}
\label{Fig4}
 \end{center}
\end{figure}
\begin{figure}[ht]
\begin{center}
  \begin{minipage}[b]{.1\textwidth}
    \begin{center}
    \centerline{ \includegraphics[width=8.5cm]{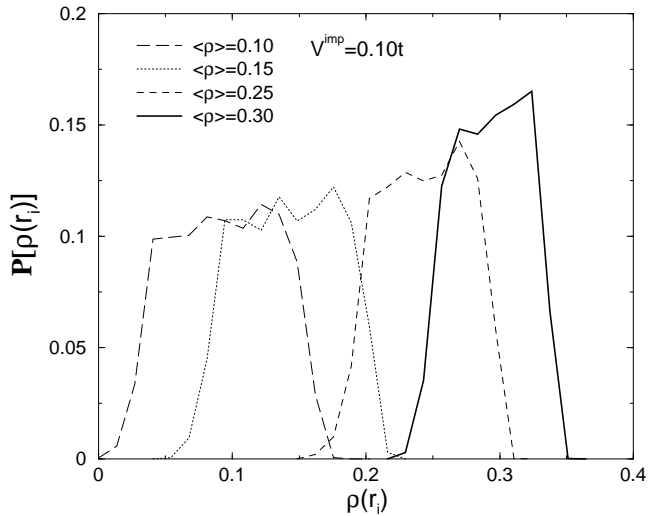}} 
    \end{center} 
  \end{minipage}
\caption{The distribution $P[\rho({\bf r}_i)]$ of local charge carriers for: 
the underdoped $\langle \rho \rangle$=0.10, $\langle \rho \rangle$=0.15, 0.25 and 0.30. 
$V^{imp}$=0.10$t$. Note that as $\langle \rho \rangle$ increases, 
$P[\rho({\bf r}_i)]$ becomes more peaked. }
\label{Fig3}
 \end{center}
\end{figure}
\begin{figure}[ht]
\begin{center}
  \begin{minipage}[b]{.1\textwidth}
    \begin{center}
    \centerline{ \includegraphics[width=8.5cm]{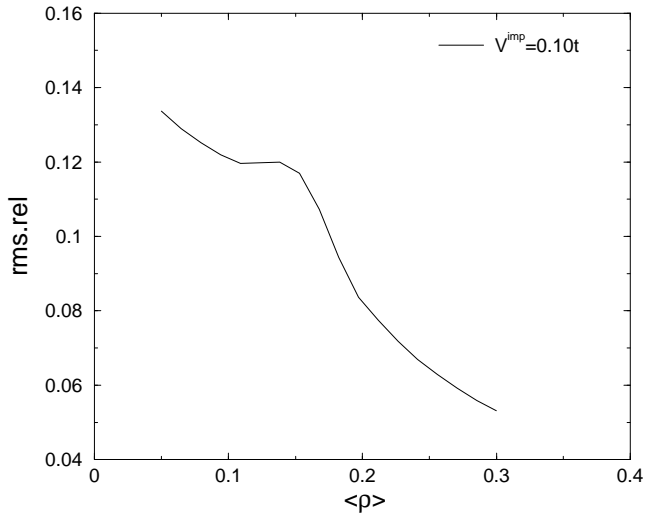}}
    \end{center} 
  \end{minipage}
\caption{The relative root mean-square deviation rms.rel for LSCO with 
$V^{imp}$=0.1$t$. It indicates that as $\langle \rho \rangle$ increase, 
rms.rel decreases showing that the local gap distributions becomes narrower. }
\label{FigA2}
 \end{center}                
\end{figure}
In this section we study the inhomogeneous case of $V^{imp}$=0.10$t$.

In Fig.\ref{Fig4} we plot the distribution $P[\Delta_0({\bf r}_i)]$ 
for some selected compounds for $V^{imp}$=0.10$t$. From the figure we see 
that as the average density $\langle \rho \rangle$ increases, 
$P[\Delta_0({\bf r}_i)]$ becomes sharp and thin. Therefore, it is clear that 
the average variation in $\Delta_0({\bf r}_i)$ decreases with increasing average doping.
In Fig.\ref{Fig3} we plot the distribution $P[\rho({\bf r}_i)]$ 
for $V^{imp}$=0.10$t$. Again we observe that increasing $\langle \rho \rangle$ 
the distributions become sharper.
\begin{figure}[ht]
\begin{center}
  \begin{minipage}[b]{.1\textwidth}
    \begin{center}
     \centerline{\includegraphics[width=4.25cm]{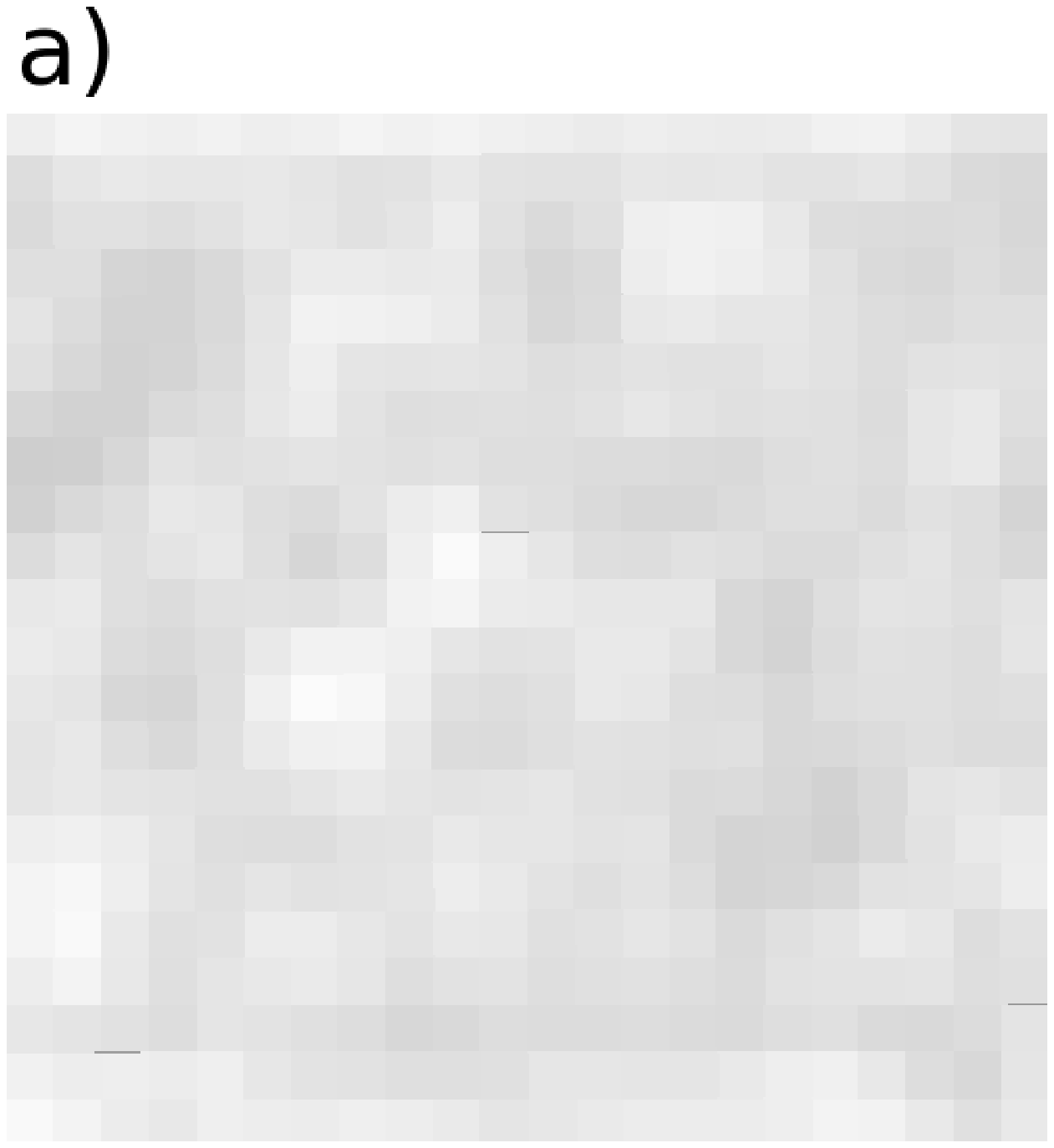}} 
    \vspace{0.45cm}
    \centerline{\includegraphics[width=4.25cm]{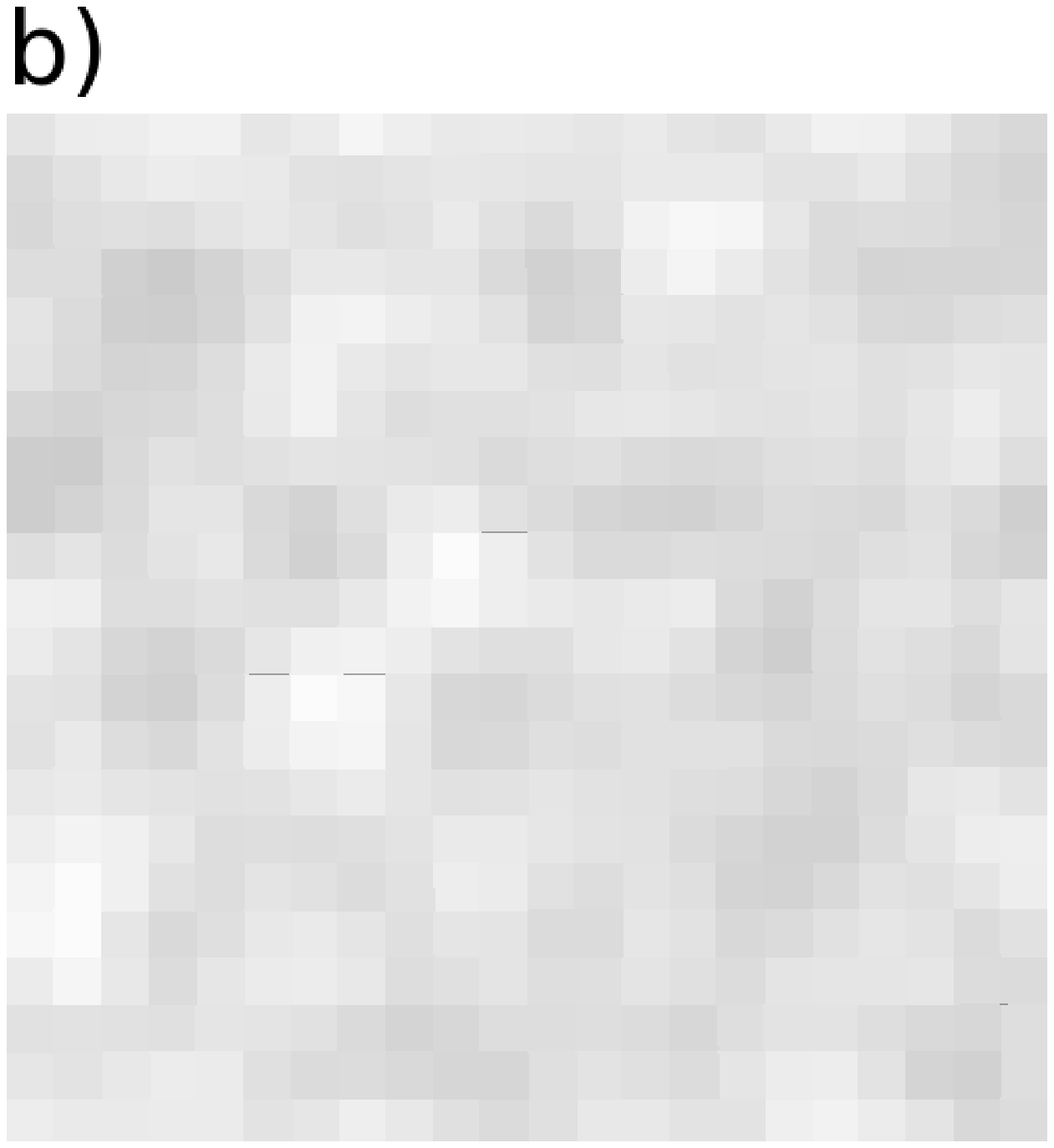}}   
    \vspace{0.45cm}
    \centerline{\includegraphics[width=4.25cm]{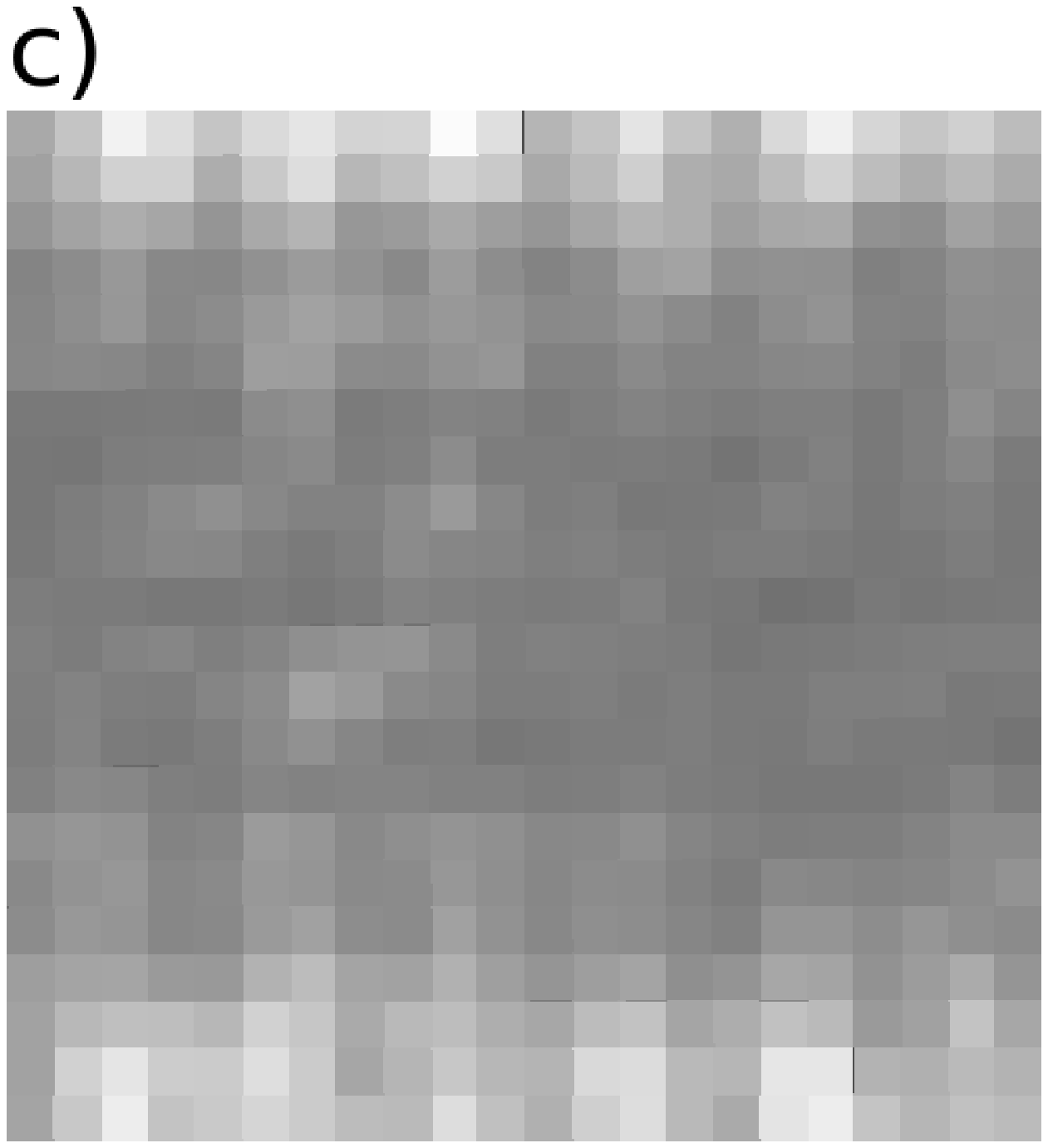}} 
    \end{center} 
  \end{minipage}
\caption{Spatial variation of $\Delta_0({\bf r}_i)$ in the 22$\times$22 lattice, 
for the mean densities $\langle \rho \rangle$=0.15 in a), 0.20 in b), and 0.30 in c).
It was considered the same realization of disorder for all compounds. In 
the figures the light regions are high $\Delta_0({\bf r}_i)$ values. Note 
that for low value of $\langle \rho \rangle$ there are many different 
color regions, indicating a more inhomogeneous medium.}
\label{Fig5}
 \end{center}
\end{figure}

To study the range of inhomogeneity, we calculate the relative root-mean-square 
deviation (rms.rel), $[\langle (\Delta\Delta_0)^2 \rangle]^{\frac{1}{2}}/\Delta_0$, 
where, by definition, $[\langle (\Delta\Delta_0)^2 \rangle]^{\frac{1}{2}}$ gives, 
approximately, a linear measure of the width of the range over which $\Delta_0$ 
is distributed.
In order to compare the rms.rel of each LSCO compound, the distributions 
$P[\Delta_0({\bf r}_i)]$ must be shifted, so they all have the same mean gap 
value $\Delta_0$, maintaining their original shape.
We can see in Fig.\ref{FigA2} that, as the average density $\langle \rho \rangle$ 
increases, rms.rel diminishes, i.e., the range over which the 
superconducting gap $\Delta_0$ is distributed also diminishes, which indicates 
again that as $\langle \rho \rangle$ goes into the overdoped region, the 
compounds become more homogeneous. 

In Fig.\ref{Fig5} we show the spatial variation map of $\Delta_0({\bf r}_i)$, 
in the 22$\times$22 lattice, 
for three different compounds for a particular realization of the random potential 
($V^{imp}$=0.10$t$): in Fig.\ref{Fig5}a we have the near-optimum 
compound ($\langle \rho \rangle$=0.15), and in Fig.\ref{Fig5}b and c, 
we have two overdoped compounds, $\langle \rho \rangle$=0.20 and 0.30, 
respectively. In the figures the light regions corresponds to high local 
superconducting gaps, and the dark regions corresponds to low local superconducting 
gaps. In Fig.\ref{Fig5}a we observe the coexistence of light regions and dark regions, 
indicating a high inhomogeneous regime. As the density increases, most of the regions 
become darker, indicating a more homogeneous medium. Therefore, we conclude again 
from the figures that, as the density $\langle \rho \rangle$ increases, the 
systems tend to become more homogeneous, in agreement with Fig.\ref{Fig4}. 
What means that a d-wave superconductor becomes more homogeneous as 
the concentration of holes or doping level increases.

\section{Conclusion}

We have calculated distributions of local $d$-wave superconducting gaps 
$\Delta_0({\bf r}_i)$ and local density of charge carriers $\rho({\bf r}_i)$ 
with possible application to the HTS La$_{2-x}$Sr$_{x}$CuO$_{4}$, 
modelled by a disordered medium. The random potential $V^{imp}$, within a BdG formalism, 
was considered as the effect of the electronic inhomogeneity produced by
the dopants. The results demonstrated that, for particular values of the 
coupling parameters and hopping integrals, the decay of $\Delta_0$ 
as $\langle \rho \rangle$ increases,  can be reproduced with this local  calculation.
We have also observed that, for the same set of parameters, the degree 
of charge inhomogeneity is much larger for a $d$ than for a $s$-wave superconductor. 
Furthermore, with all parameters fixed but the average density, 
the underdoped compounds have a clear tendency to be more inhomogeneous 
than the overdoped ones, which is in agreement with experimental findings~\cite{Muller,Singer,Buzin00}. 
As shown in Fig.\ref{Fig5}, the spatial variation of local superconducting gaps 
exhibits regions of high $\Delta_0({\bf r}_i)$ enclosed by 
low $\Delta_0({\bf r}_i)$ regions, which favours the possibility that the 
superconducting phase is reached by the percolation of many superconducting
patches or stripes~\cite{Evandro3,Evandro2,Ovchinnikov}.

We acknowledge partial support from the CNPq and CNPq/Faperj Pronex E26/171.168/2003.


\end{document}